\documentclass[twocolumn,prb,amssymb,showpacs]{revtex4}
\usepackage{graphicx}

\begin{document}

\title{Electron-electron interaction
at decreasing $k_Fl$ }

\author{G.~M.~Minkov}
\email{Grigori.Minkov@usu.ru}
\author{O.~E.~Rut}
\author{A.~V.~Germanenko}
\author{A.~A.~Sherstobitov}
\affiliation{Institute of Physics and Applied Mathematics, Ural State
University, 620083 Ekaterinburg, Russia}

\author{V.~I.~Shashkin, O.~I.~Khrykin}
\affiliation{Institute of Physics of Microstructures of RSA, 603600
Nizhni Novgorod, Russia}

\author{B.~N.~Zvonkov}
\affiliation{Physical-Technical Research Institute, University of
Nizhni Novgorod, 603600 Nizhni Novgorod, Russia}

\date{\today}

\begin{abstract}
The contribution of the electron-electron interaction to conductivity
is analyzed step by step in gated GaAs/InGaAs/GaAs heterostructures
with different starting disorder. We demonstrate that the diffusion
theory works down to $k_F l\simeq 1.5-2$, where $k_F$ is the Fermi
quasimomentum, $l$ is the mean free paths. It is shown that the e-e
interaction gives smaller contribution to the conductivity than the
interference independent of the starting disorder and its role rapidly
decreases with $k_Fl$ decrease.
\end{abstract}

\maketitle

The quantum corrections to the conductivity in disordered metals and
doped  semiconductors are intensively studied since 1980. Two
mechanisms lead to these corrections: (i)  the interference of the
electron waves propagating in opposite directions along closed paths;
(ii)  electron-electron (e-e) interaction. The absolute value of these
corrections increases with decreasing  temperature and/or increasing
disorder and they determine in large part the low temperature transport
in 2D systems.

The interference correction $\delta\sigma^{WL}$ is proportional to
$-\ln(\tau_\phi/\tau)$, where $\tau_\phi$ and $\tau$ are the phase and
momentum relaxation time, respectively, $\tau_\phi \propto T^{-p}$, $p
\simeq 1$. The correction due to e-e interaction $\delta\sigma^{ee}$ is
proportional to $-\ln[\hbar/(k_B T \tau)]$. \cite{Altshuler} It
immediately  follows that at increasing disorder, i.e. at decreasing
$\tau$, both corrections have to be enhanced in absolute value and can
become comparable with the Drude conductivity. In this case the low
temperature conductivity will be significantly less than the Drude
conductivity and strong temperature dependence of the conductivity has
to appear. On further disorder increasing the transition to the hopping
conductivity has to occur.

Conventional theories of the quantum corrections both in the diffusive
$k_B T\tau/\hbar \ll 1$ \cite{Altshuler} and in the ballistic
\cite{Aleiner, Gornyi} regimes was developed for the case $k_F l \gg
1$, where $k_F$ and $l$ are the Fermi quasimomentum and the classical
mean free path, respectively. Under this condition the quantum
corrections to the conductivity are small in magnitude compared with
the Drude conductivity $\sigma_0=\pi k_F l G_0$ with
$G_0=e^2/(2\pi^2\hbar)$ at any accessible temperature. At decreasing
$k_F l $ the relative values of the quantum corrections are enhanced
and the question is how the values of these corrections and their ratio
changes when $k_Fl$ tends to 1.

In our previous paper \cite{ourWLSL} we have shown that the
contribution of the e-e interaction to the conductivity decreases at
decreasing $k_Fl$. In the present paper we study $k_F$ dependence of
the contribution to the conductivity due to e-e interaction in
structures distinguished by a starting disorder. We demonstrate that
(i) the diffusion theory works down to $k_F l\simeq 1.5-2$ (ii) the e-e
interaction gives smaller contribution to the conductivity than the
interference independent of the starting disorder and its role rapidly
decreases with $k_Fl$ decrease.

Two types of the heterostructures with 80\AA -In$_{0.2}$ Ga$_{0.8}$As
single quantum well in GaAs were investigated.  Structures 1 and 2 with
relatively high starting disorder had Si $\delta$ doping layer in the
center of the quantum well. The electron density $n$ and mobility $\mu$
in these structures were the following: $n=1.45\times10^{16}$~m$^{-2}$
and $\mu=0.19$~ m$^2$/Vs in structure 1, $n=0.89\times10^{16}$~m$^{-2}$
and $\mu=0.23$~ m$^2$/Vs in structure 2. Structures 3 and 4 had lower
starting disorder because the doping $\delta$ layers were disposed on
each side of the quantum well and were separated from it by the 60~\AA\
spacer of undoped GaAs. The values of $n$ and $\mu$ were:
$n=5.1\times10^{15}$~m$^{-2}$ and $\mu=13.0$~ m$^2$/Vs in structure 3,
$n=2.3\times10^{15}$~m$^{-2}$ and $\mu=13.9$~m$^2$/Vs in structure 4.
The thickness of undoped GaAs cap layer was 3000 \AA\ for all
structures. The samples were mesa etched into standard Hall bars and
then an Al gate electrode was deposited by thermal evaporation onto the
cap layer through a mask. Varying the gate voltage $V_g$ from $0.0$ to
$-3..-4$ V we decreased the electron density in the quantum well and
changed $k_F l$ from its maximal value ($9-30$ for different
structures) down to $\simeq 1$.

\begin{figure}
\includegraphics[width=7cm,clip=true]{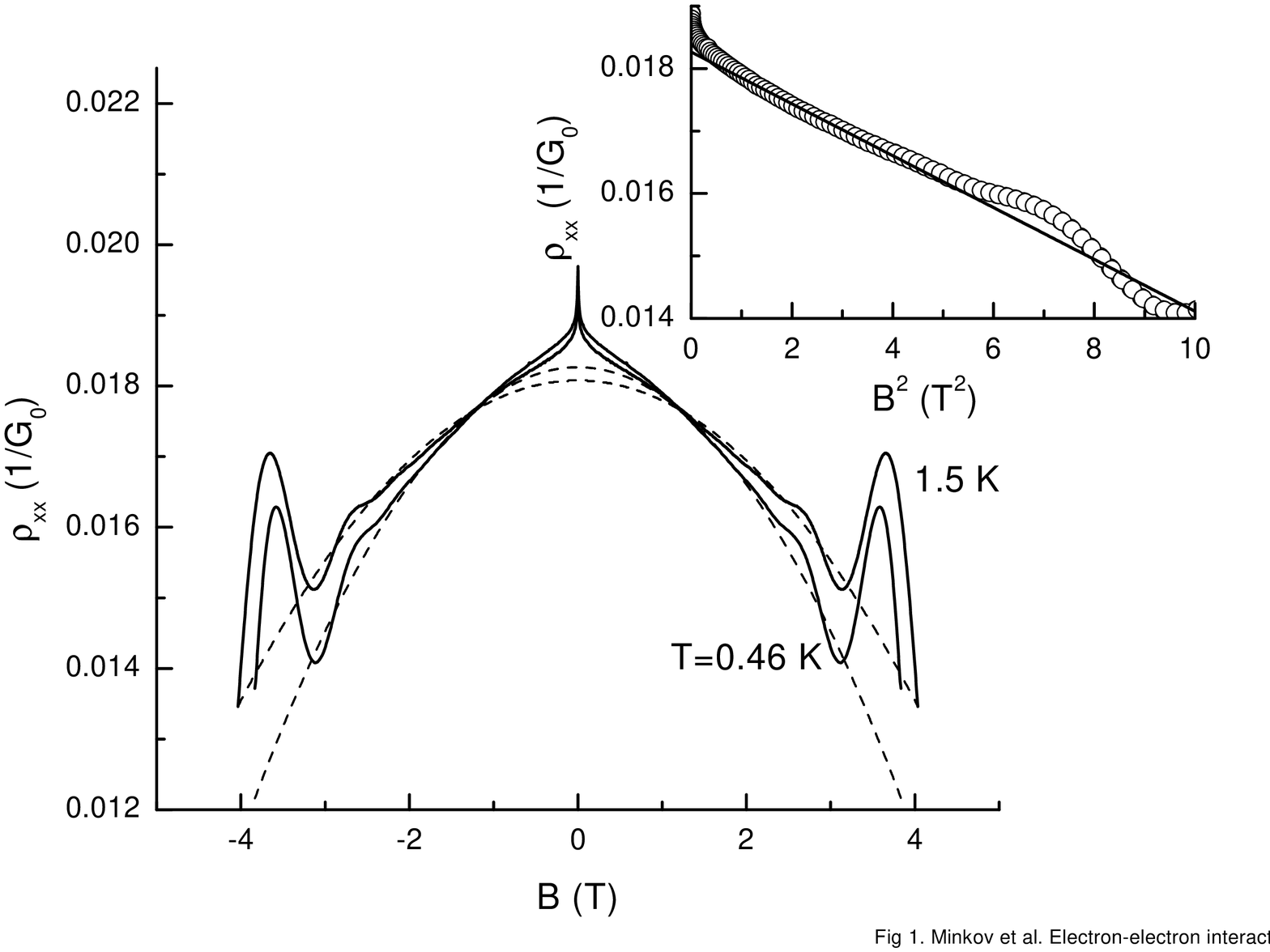}
 \caption{The magnetic field dependence of $\protect\rho_{xx}$
for structure 3 at $k_F l=17.9$. Solid curves are the experimental
data, dashed lines are Eq.~(\protect\ref{eq6}) with parameters
corresponding to the best fit carried out in the range from $\pm 1$ to
$\pm 3.2$~T which gives $K_{ee}$=0.35 and 0.34 for $T=0.46$~K and
$1.5$~K, respectively. Inset shows $\rho_{xx}$ as a function of $B^2$
for $T=0.46$~K. } \label{fig1}
\end{figure}

Figure~\ref{fig1} shows the experimental magnetic field dependences of
$\rho_{xx}$ measured at two temperatures for one of the structure when
$k_F l=17.9$. Two different magnetic field ranges are evident: the
range of sharp dependence of $\rho_{xx}$ at low field $B\leq 0.3$~T,
and the range of moderate dependence which is close to parabolic one at
higher field. The feature is the fact that $\rho_{xx}$-vs-$B$ curves
for different temperatures cross each other at magnetic field
$B_{cr}=1.1$~T which is close to $\mu^{-1}$.

The low-magnetic-field negative magnetoresistance is caused by the
suppression of the interference quantum correction. The characteristic
magnetic field scale for this effect is so called transport magnetic
field $B_{tr}=\hbar /(2el^2)$ which is equal to approximately 0.03~T in
the given case. So the interference quantum correction can be easily
separated due to its sharp specific magnetic field dependence.

The parabolic negative magnetoresistance in higher magnetic field
results from the contribution of the e-e interaction.\cite{Paalen}
Since this effect is more pronounced in relatively high magnetic fields
of order $\mu^{-1}$ where other classical mechanisms of both positive
and negative\cite{Bask,Bob,Dyak,Polyak} magnetoresistance can be
efficient it is significantly more complicated to analyze it
quantitatively.

To separate the electron-electron contribution to the conductivity we
have analyzed the data by the same way as in our previous
papers.\cite{ourPr,ourWLSL} Specific feature of the e-e interaction is
the fact that it contributes to $\sigma_{xx}$ only and this
contribution does not depend on the magnetic field until
$g\mu_BB/k_BT<1$:

\begin{eqnarray}
\delta \sigma^{ee}_{xx}&=&-\left(1+\frac{3}{4}\lambda\right)G_0\ln\frac{\hbar}{k_BT\tau} \label{eq1a}\\
\delta \sigma^{ee}_{xy}&=&0.\label{eq1b}
\end{eqnarray}
Here, $\lambda$ has been calculated in Ref.~\onlinecite{Fin}, it is a
function of $k_F/K$ with $K $ as the screening parameter which for 2D
case is equal to $2/a_B$, where $a_B$ is the effective Bohr radius.
Eq.~(\ref{eq1a}) is valid in the diffusion regime when $k_BT \tau/\hbar
\ll 1$. Theory for ballistic and intermediate regime was developed in
Ref.~\onlinecite{Aleiner} for short range scattering potential and in
Ref.~\onlinecite{Gornyi} for long range potential. In our case $k_BT
\tau/\hbar <0.25$ under all conditions, therefor we believe the
diffusion approximation is valid.

\begin{figure}
\includegraphics[width=\linewidth,clip=true]{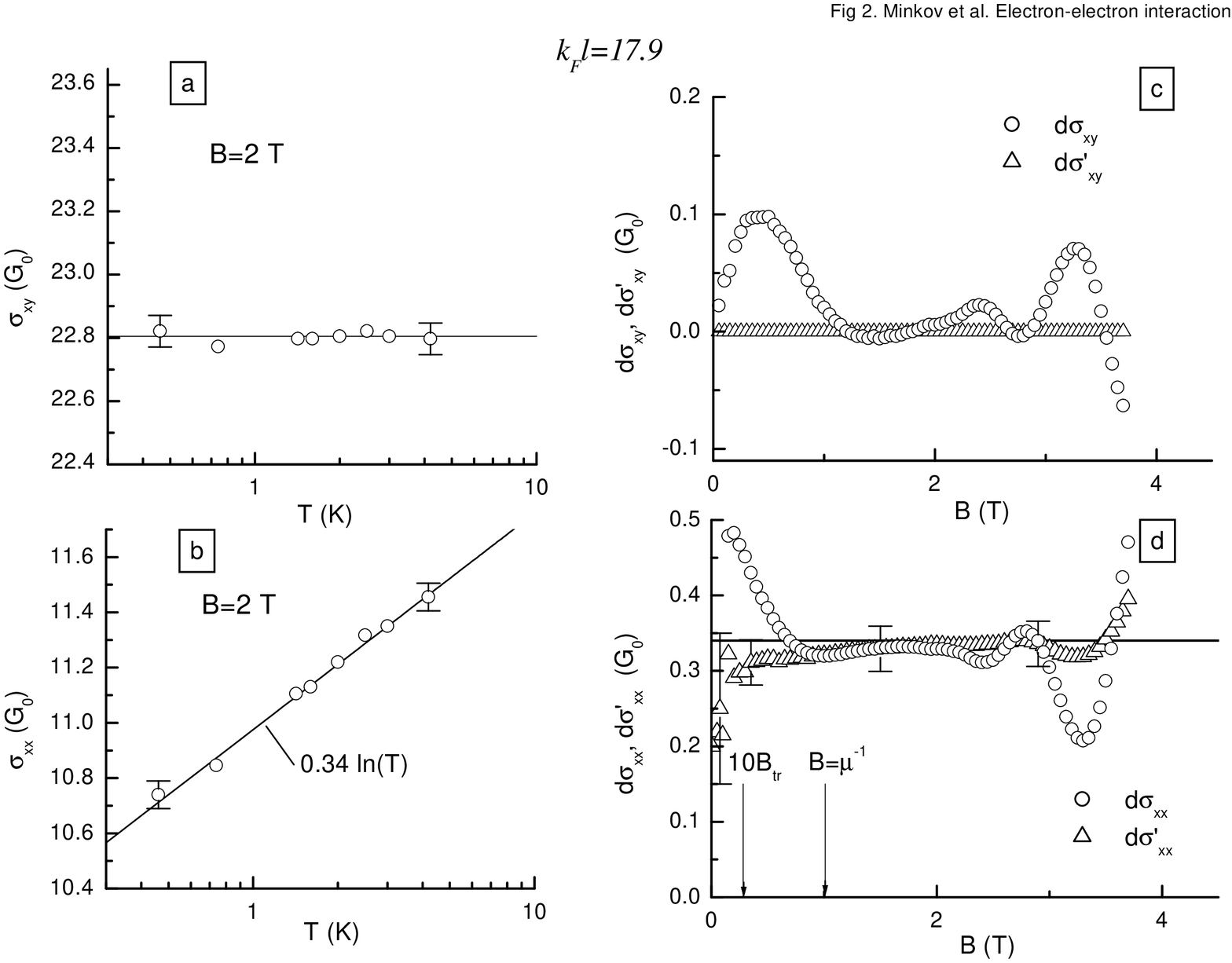}
 \caption{The temperature dependence of $\protect\sigma_{xx}$ (a) and $\protect%
\sigma_{xy}$ (b) for $B=2$~T. The magnetic filed dependence of
$d\sigma_{xy}$, $d\sigma_{xy}'$ (c) and $d\sigma_{xx}$, $d\sigma_{xx}'$
(d) obtained with $T_1=4.2$~K and $T_2=0.46$~K. Straight line in (d)
corresponds to $K_{ee}=0.34$ found from the temperature dependence of
$\sigma_{xx}$ at $B=2$~T depicted in (b). Structure 3, $k_Fl=17.9$. }
\label{fig2}
\end{figure}

Thus, at those magnetic fields where the interference correction to the
conductivity has been fully suppressed, the behavior of the
conductivity tensor components  corresponding to Eqs.~(\ref{eq1a}) and
(\ref{eq1b}) has to be observed. When it is the case one can find the
value of prefactor $K_{ee}=-(1+3/4\lambda)$ in Eq.~(\ref{eq1a}) and so
the contribution of the e-e interaction.  Just such behavior of
$\sigma_{xx}$ and $\sigma_{xy}$ is observed for both types of
structures when $k_Fl\gg 1$. As an example the experimental temperature
dependences of $\sigma_{xx}$ and $\sigma_{xy}$ taken at B=2~T are
presented in Figs.~\ref{fig2}(a), \ref{fig2}(b) for the structure 3
when $k_Fl \simeq 18$. On can see that the value of $\sigma_{xx}$
really logarithmically decreases when the temperature decreases whereas
$\sigma_{xy}$ is temperature independent.

To show the magnetic field range in which such a behavior of
$\sigma_{xx}$ and $\sigma_{xy}$ with temperature takes place, the
differences $d \sigma_{xy}(B) = [\sigma_{xy}(B,T_1)-
\sigma_{xy}(B,T_2)]/\ln(T_1/T_2)$ and $d\sigma_{xx}(B) =
[\sigma_{xx}(B,T_1)- \sigma_{xx}(B,T_2)]/\ln(T_1/T_2)$ as a function of
magnetic field are plotted in Figs.~\ref{fig2}(c), \ref{fig2}(d) by
circles. In the situation when only the e-e interaction contributes to
the conductivity, $d \sigma_{xy}(B)$ and $d \sigma_{xx}(B)$ have to be
independent of the magnetic field and must be equal to zero and
$K_{ee}$, respectively. One can see that $d\sigma_{xy}(B)$, really, ten
times less than $d\sigma_{xx}(B)$ within magnetic field range from
0.8~T to 3~T. Therewith $d\sigma_{xx}(B)$ is close to constant which in
its turn corresponds to the value of $K_{ee}$ found from the
temperature dependence of $\sigma_{xx}$ [Fig.~\ref{fig2}(b)].

\begin{figure}
\includegraphics[width=\linewidth,clip=true]{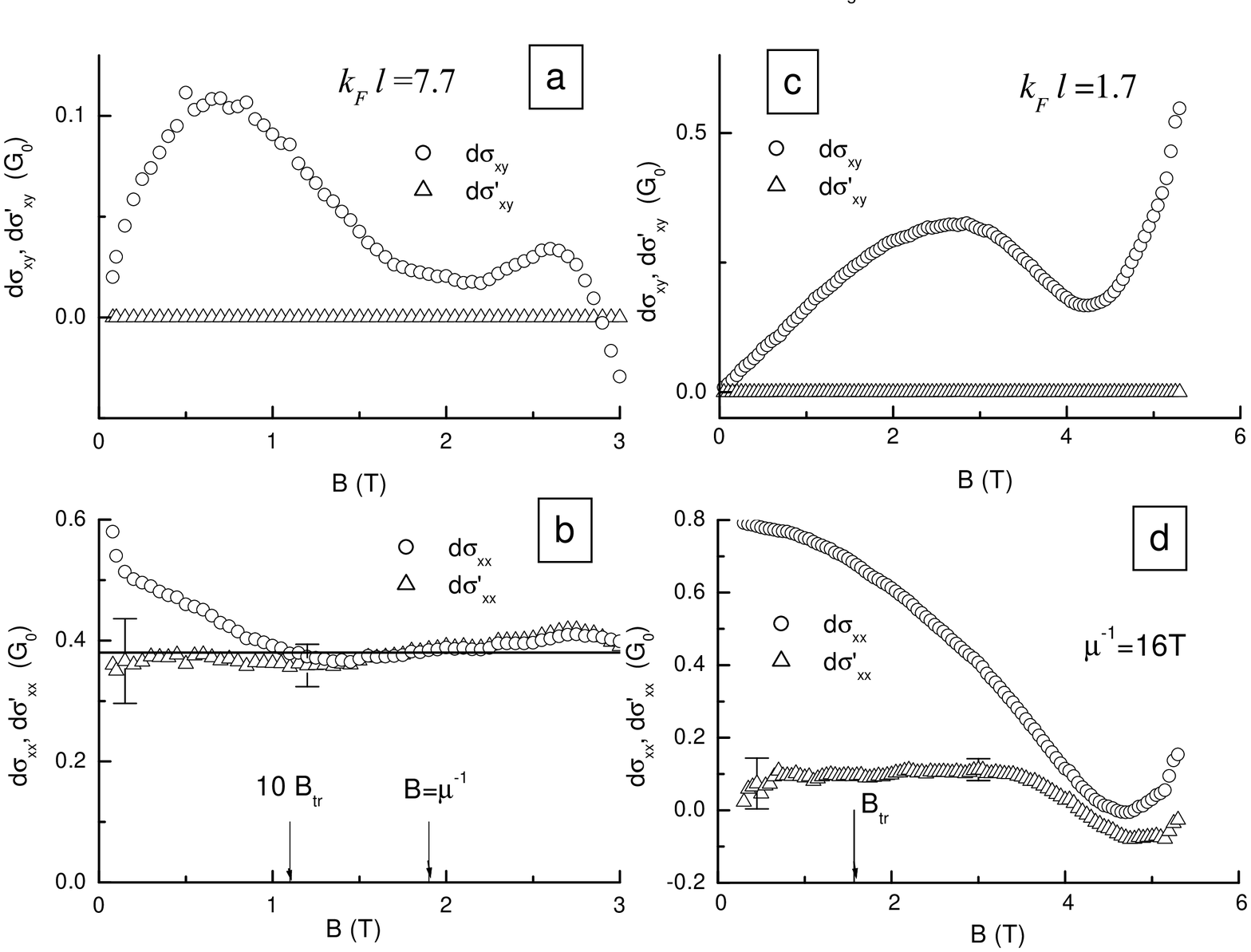}
\caption{The magnetic filed dependence of $d\sigma_{xy}$,
$d\sigma_{xy}'$ (a,c) and $d\sigma_{xx}$, $d\sigma_{xx}'$ (b,d) for
$k_Fl=7.7$ (a,b) and $k_Fl=1.7$ (c,d) for structure 3, $T_1=4.2$~K,
$T_2=0.46$~K. Straight line in (b) corresponds to $K_{ee}=0.38$ found
from the temperature dependence of $\sigma_{xx}$ for $B=2$~T. }
\label{fig3}
\end{figure}

We consider what sets the limits on the magnetic field range in which
$d\sigma_{xy}\ll d\sigma_{xx}$ and $d\sigma_{xx}/G_0\simeq K_{ee}$. The
interference correction does it on the low-magnetic-field side. Note,
that this correction leads to appreciable changes in $d\sigma_{xx}(B)$,
which is comparable with the contribution due to e-e interaction up to
$(10-20) B_{tr}$. On the high-magnetic-field side the limitation is
caused by the Shubnikov-de Haas oscillations which appear when
$B>(1-1.5)\mu^{-1}$. Thus, $d\sigma_{xx}(B)$ is constant and
$d\sigma_{xy}\ll d\sigma_{xx}$ within the magnetic field range
$(10-20)B_{tr}<B<(1-1.5)\mu^{-1}$. The ratio $\mu^{-1}/B_{tr}$ is equal
to $2k_Fl$ therefor the magnetic field range where $d\sigma_{xy}\ll
d\sigma_{xx}$ fast narrows at decreasing $k_Fl$. So, for $k_F l=17.9$
the range where $d\sigma_{xy}< 0.1 d\sigma_{xx}$ is $0.8-3$~T, for
$k_Fl=7.7$ it is $1.5-3$~T [Figs.~\ref{fig3}(a), \ref{fig3}(b)], and
finally for $k_Fl\simeq 1.7$ \footnote{The value of the Drude
conductivity was found as
$\sigma_0=\sigma(T)-\delta\sigma^{ee}-\delta\sigma^{WL}$. This value
coincides with an accuracy of $G_0$ with the conductivity at $T=40$~K.}
such range is quite absent [Figs.~\ref{fig3}(c), \ref{fig3}(d)]. Thus,
the absence of the range of magnetic field, in which the interference
correction is significantly less than the e-e correction, makes it
impossible to determine $K_{ee}$ for low $k_Fl$ values.

Let us attempt to extract the interference contribution from
$\sigma_{xx}$. We will use the fact that the interference gives the
contribution to back scattering and hence to the transport relaxation
time. \cite{Altshuler} Thus, the interference corrections to both
components of the conductivity tensor are nonzero
\begin{eqnarray}
\sigma_{xx}(B,T)&=&\frac{en\mu}{1+\mu^2B^2}+\delta\sigma_{xx}^{WL}(B,T)+
\delta\sigma_{xx}^{ee}(T) \label{eq3}\\
\sigma_{xy}(B,T)&=&\frac{en\mu^2B}{1+\mu^2B^2}+\delta\sigma_{xy}^{WL}(B,T)
\label{eq4}.
\end{eqnarray}
If $\delta\sigma_{xy}^{WL}\ll\sigma_{xy}$ and
$\delta\sigma_{xx}^{WL}\ll \sigma_{xx}$, the following simple
relationship is valid
\begin{equation}
\frac{\delta\sigma_{xy}^{WL}}{\sigma_{xy}}=2
\frac{\delta\sigma_{xx}^{WL}}{\sigma_{xx}}. \label{eq5}
\end{equation}
Thus, we can determine  $\delta\sigma_{xy}^{WL}(B,T_1)-
\delta\sigma_{xy}^{WL}(B,T_2)$ as difference between the experimental
curves $\sigma_{xy}(B)$ taken at $T_1$ and $T_2$, calculate
$\delta\sigma_{xx}^{WL}(B,T_1)- \delta\sigma_{xx}^{WL}(B,T_2)$ from
Eq.~(\ref{eq5}), and then extract this difference from the experimental
$\sigma_{xx}(B,T_1)-\sigma_{xx}(B,T_2)$ curve. Dividing the results by
$\ln(T_1/T_2)$ we obtain $d\sigma_{xx}'(B)$ which does not contain the
interference contribution and has to be equal to $K_{ee}$, in
principle, starting from zero magnetic field.

The procedure described has been checked by analyzing the results for
structure 3 at high value of $k_F l$ presented above. The results are
shown in Figs.~\ref{fig2}(c), \ref{fig2}(d) and Fig.~\ref{fig3}(a),
\ref{fig3}(b)  by triangles. Self-evident $d\sigma_{xy}'(B)$ vanishes,
whereas $d\sigma_{xx}'(B)$ becomes constant starting from the low
magnetic field and is equal to $K_{ee}$ obtained from the temperature
dependence of $\sigma_{xx}$ at high magnetic field.

Now we are in position to analyze the results for low $k_Fl$ value. As
mentioned above there was no magnetic field range where $d\sigma_{xy}$
was much smaller than $d\sigma_{xx}$, and $d\sigma_{xx}$ did not depend
on the magnetic filed for $k_Fl\simeq 1.7$. After extraction of the
interference contribution we have obtained the wide range of magnetic
field from 0.5 to 3.5~T where $d\sigma_{xx}'\simeq$~const
[Figs.~\ref{fig3}(c), \ref{fig3}(d)]. This aloows us to believe that
$d\sigma_{xx}'/G_0$ gives the value of $K_{ee}$. Strictly speaking, the
interference corrections $\delta\sigma_{xy}^{WL}$ and
$\delta\sigma_{xx}^{WL}$ can be comparable in magnitude with
$\sigma_{xy}$ and $\sigma_{xx}$ respectively if the parameter $k_Fl$ is
small enough. In this case the relation between the interference
corrections is more cumbersome than Eq.~(\ref{eq5}) and we do not write
it out. We note only that the use of the rigorous formula gives the
result which lies within an error indicated in Fig.~\ref{fig3}(d).

\begin{figure}
\includegraphics[width=\linewidth,clip=true]{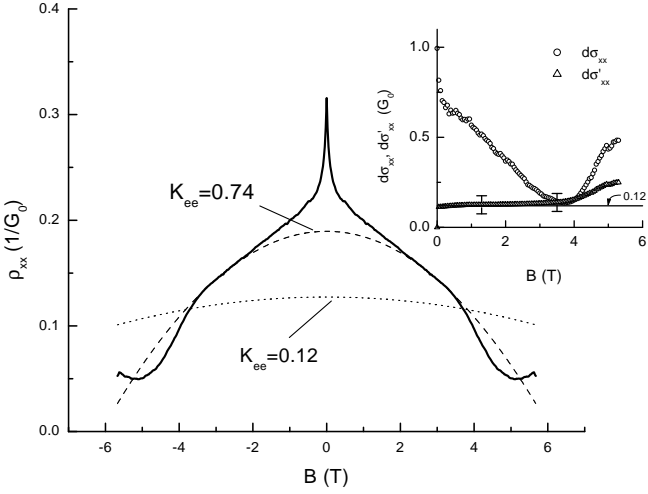}
\caption{The magnetic filed dependence of $\rho_{xx}$ for $k_Fl=2.8$
(structure 3, $n=2.55\times 10^{15}$~m$^{-2}$), $T=0.46$~K. Solid line
is the experimental data. Dotted line is Eq.~(\ref{eq6}) with correct
value of $\delta\sigma_{ee}$ corresponding to $K_{ee}=0.12$, dashed
line is the best fit by Eq. (\ref{eq6}) carried our in the range from
$\pm 2$ to $\pm5.8$~T, which gives however wrong value of
$K_{ee}=0.74$. Inset is the magnetic field dependence of
$d\sigma_{xx}$, $d\sigma_{xx}'$ which illustrates obtaining correct
value of $K_{ee}=0.12$.} \label{fig4}
\end{figure}

Before discussion of the final results let us turn to the procedure of
determination of $K_{ee}$, used in Refs.~\onlinecite{Paalen,Pourier,
Nina, Proskuryakov}.  The contribution of e-e interaction to the
conductivity was determined from the negative parabolic
magnetoresistance which directly follows from (\ref{eq1a}) and
(\ref{eq1b}) for low $\delta\sigma_{xx}^{ee}$ value
\begin{equation}
\rho_{xx}(B,T)  \simeq
\frac{1}{\sigma_0}-\frac{1}{\sigma_0^2}\left(1-\mu^2
B^2\right)\delta\sigma_{xx}^{ee}(T). \label{eq6}
\end{equation}
This method can be applied at $k_Fl\gg1$ when there is the wide
magnetic field range where the contribution due to the interference is
significantly less than due to e-e interaction. As is seen from
Fig.~\ref{fig1} it gives the value of $K_{ee}$ close to that obtained
from the temperature dependence of $\sigma_{xx}$ [Fig.~\ref{fig2}(b)].
At low $k_Fl$ values the magnetic field dependence of $\rho_{xx}$ can
be also described by the parabola as shown in Fig.~\ref{fig4}. However,
the parameter of the e-e interaction $K_{ee}$ found from the fit can
dramatically differ from the correct value. It naturally follows from
the fact that for low $k_Fl$ the interference correction significantly
influences the magnetic field dependence of $\rho_{xx}$ in wide range
of magnetic fields.

\begin{figure}
\includegraphics[width=\linewidth,clip=true]{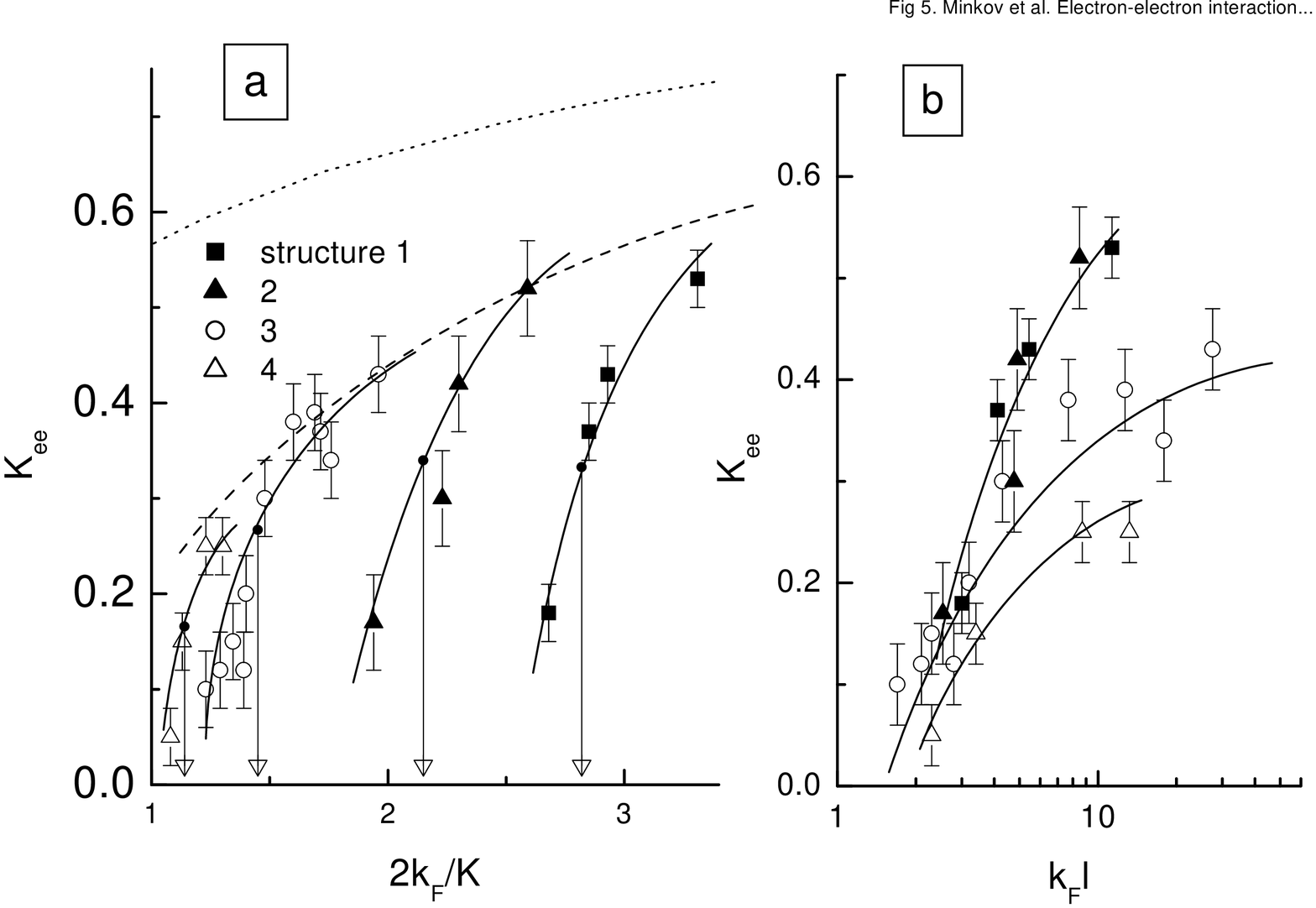}
\caption{The value of $K_{ee}$ as a function of $2k_F/K$ (a) and $k_Fl$
(b). Symbols are the experimental data. Dotted line is result from
Ref.~\protect\onlinecite{Fin}, dashed line is drown through the dots
with highest values of $k_F l$, solid lines are provided as a guide for
the eye. Arrows in (a) indicate the $2k_F/K$ values at which $k_Fl=4$
for different structures.}
 \label{fig5}
\end{figure}

Let us return now to our results. The $K_{ee}$-versus-$k_F$ dependence
for all the structures investigated are presented in Fig.~\ref{fig5}(a)
together with theoretical curve calculated according to
Ref.~\onlinecite{Fin}. Consider first the points with highest $k_F l$
(i.e. with highest $k_F$) for each structure. One can see that they
fall on the one smooth curve (dashed line in the figure) which lies
somewhat below the theoretical one. The deviation is stronger for
structures 3 and 4 with lower disorder in which parameter $k_B
T\tau/\hbar$ is about $0.25$ for $T=4.2$~K. Probably, this value is not
sufficiently small and the diffusion approximation $k_B T\tau/\hbar\ll
1$ is rather crude.\cite{Aleiner, Gornyi}

Seemingly, at decreasing $k_F$ with gate voltage the experimental
points for every structure have to move left along dotted line. However
as clearly seen they sharply deviate down. This results from the
decrease of $k_F l$ with $k_F$ decrease. To illustrate the above we
present $k_F l$ dependence of $K_{ee}$ in Fig.~\ref{fig5}(b). Thus,
$K_{ee}$ decreases with decreasing $k_Fl$ for all the structures with
different starting disorder and the lower is the value of $k_Fl$, the
stronger is the deviation from the theory [see Fig.~\ref{fig5}(a)]. It
is not surprising because the theory was developed for the case
$k_Fl\gg 1$. Besides, the scattering by the short-range scattering
potential was taken into account only whereas the role of long-range
scattering potential is enhanced at decrease of the electron density
with the gate voltage.

\begin{figure}
\includegraphics[width=\linewidth,clip=true]{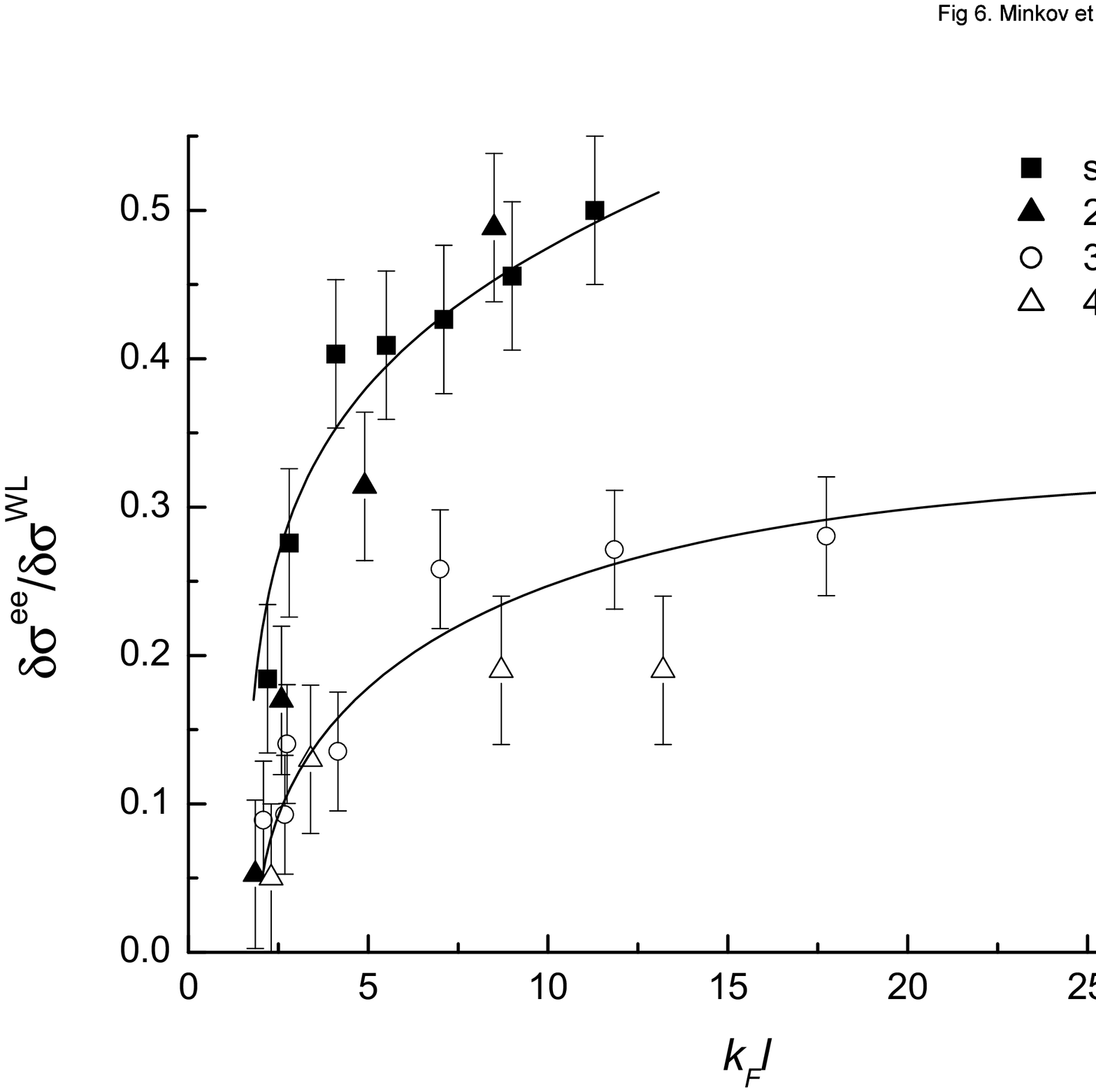}
\caption{The $\delta\sigma^{ee}$ to $\delta\sigma^{WL}$ ratio as a
function of $k_Fl$, $T=0.46$~K. Symbols are experimental data. Solid
lines are a guide for the eye. }
 \label{fig6}
\end{figure}

Next, we compare the value of the correction to the conductivity due to
the e-e interaction with that due to the interference. The value of the
interference correction was found as $-G_0 \ln(\tau_\phi/\tau)$ with
$\tau_\phi$ obtained from the low-magnetic-field negative
magnetoresistance. \cite{Hik,schm,ourPr} The $\delta\sigma^{ee}$ to
$\delta\sigma^{WL}$ ratio for $T=0.46$~K as a function of $k_Fl$ for
the structures investigated is plotted in Fig.~\ref{fig6}. One can see:
(i) the contribution due to the e-e interaction is always smaller than
that due to the interference; (ii) the relative contribution of e-e
interaction is somewhat larger in the structures 1 and 2 with doped
well, i.e., with higher disorder; (iii) the relative role of the e-e
interaction rapidly decreases with decreasing $k_Fl$ for both types of
structure independent of the starting disorder.

Thus, the main correction to the conductivity in our structures comes
from the interference rather than from the e-e interaction. Just the
interference correction can be comparable in magnitude with the Drude
conductivity at low $k_F l$ and lead, thus, to the strong temperature
dependence of the conductivity in this case.

This conclusion is opposite to that obtained for thin metal films. The
tunneling and transport investigations reveal that namely the e-e
interaction is responsible for strong decrease of the low temperature
conductivity of the metal films (see for example
Ref.~\onlinecite{metal}). The possible reason for this difference is
the fact that in contrast to the structures investigated the strong
spin-orbit interaction in metal suppresses the interference correction
and makes thus the e-e interaction correction most important.

In summary, the contribution of the electron-electron interaction to
the conductivity of 2D electron gas has been studied in gated
GaAs/InGaAs structures with different starting disorder. To obtain the
reliable data for low values of $k_F l$, the method for separation of
the e-e contribution has been proposed. It has been shown that
independent of the starting disorder the value of $-(1+3/4\lambda)$ is
close to the theoretical one for high value of $k_F l $ and exhibits a
dramatic decrease with lowering $k_F l $. We have found that the e-e
interaction gives smaller contribution to the conductivity than the
interference and its role rapidly decreases with decreasing $k_Fl$.

\subsection*{Acknowledgments}
We thank Igor Gornyi for useful discussion. This work was supported in
part by the RFBR through Grants No. 00-02-16215, No. 01-02-16441 and
No. 01-02-17003, the INTAS through Grant No. 1B290, the Program {\it
University of Russia} through Grants No.~UR.06.01.002, the CRDF through
Grant No. REC-005.


\begin{thebibliography}{}
 \bibitem{Altshuler} B. L. Altshuler, and A. G. Aronov, in {\em Electron-Electron
Interaction in Disordered Systems}, edited by A.~L.~Efros and
M.~Pollak, (North Holland, Amsterdam, 1985) p.1.
\bibitem{Aleiner} Gabor Zala, B. N. Narozhny, and I. L. Aleiner, Phys. Rev. B {\bf 64}, 214204
(2001), Phys. Rev. B {\bf 64}, 201201 (2001).
\bibitem{Gornyi} I.V. Gornyi  and A.D. Mirlin cond-mat/0207557.
\bibitem{ourWLSL} G. M. Minkov, O. E. Rut, A. V. Germanenko,
A. A. Sherstobitov, B. N. Zvonkov, E. A. Uskova, and A. A. Birukov,
Phys. Rev. B {\bf 65}, 235322 (2002).
\bibitem{Paalen} M. A. Paalanen, D. C. Tsui, and J. C. M. Hwang, Phys. Rev. Letters {\bf
51}, 2226 (1983).
\bibitem{Bask} E.M. Baskin, L.N. Magarill, and M.V. Entin, Sov. Phys. JETP {\bf 48}, 365
(1978).
\bibitem{Bob} A. V. Bobylev, Frank A. Maao, Alex Hansen, and E. H.
Hauge, Phys. Rev. Lett. {\bf 75}, 197 (1995).
\bibitem{Dyak}Alexander Dmitriev, Michel Dyakonov, and Remi Jullien, Phys. Rev. B {\bf 64}, 233321
(2001).
\bibitem{Polyak}A. D. Mirlin, D. G. Polyakov, F. Evers, and P. W\"{o}lfle
Phys. Rev. Lett. {\bf 87}, 126805 (2001).
\bibitem{ourPr} G. M. Minkov, O. E. Rut, A. V. Germanenko, A. A. Sherstobitov, V. I.
Shashkin, O. I. Khrykin, and V. M. Daniltsev Phys. Rev. B {\bf 64},
235327 (2001).
\bibitem{Fin} A.~M.~Finkelstein, Zh. Eksp. Teor. Fiz. {\bf 84}, 168 (1983)
[Sov. Phys. JETF {\bf 57}, 97 (1983)].
\bibitem{Pourier} W. Poirier, D. Mailly, and M. Sanquer, Phys. Rev. B {\bf 57}, 3710
(1998).
\bibitem{Nina} Yu. G. Arapov, G. I. Harus, O. A. Kuznetsov, V. N. Neverov, and
N. G. Shelushinina, Fiz. Tekn. Poluprov. {\bf 33}, 1073 (1999)
[Semiconductors {\bf 33}, 978 (1999)].
\bibitem{Proskuryakov} L. Li, Y.Y. Proskuryakov, A.K. Savchenko, E.H. Linfield, and D.A.
Ritchie, cond-mat/0207662.
 \bibitem{Hik} S.~Hikami, A.~Larkin and Y.~Nagaoka, Prog. Theor. Phys. {\bf
63}, 707 (1980).
\bibitem{schm}  H.-P.~Wittman and A.~Schmid, J. Low. Temp. Phys. {\bf 69},
131 (1987).
\bibitem{metal} G. Hertel, D. J. Bishop, E. G. Spencer, J. M. Rowell, and R. C.
Dynes, Phys. Rev. Lett. {\bf 50}, 743 (1983); V. Yu. Butko, J. F.
DiTusa, and P. W. Adams, Phys. Rev. Lett. {\bf 84}, 1543 (2000)

\end{thebibliography}
\end{document}